\documentclass[default]{aastex701}

\usepackage{soul, xcolor}
\usepackage{amsmath}

\usepackage{adjustbox}
\usepackage{graphicx}

\begin{document}

\title{Disruption limits around triaxial bodies and implications for ring-bearing minor bodies}

\author[0000-0001-5138-230X]{Gustavo~Madeira}
\affiliation{Observat\'orio Nacional/MCTI, R. General Jos\'e Cristino 77, CEP 20921-400, RJ, Brazil}
\email[show]{madeira@on.br}  

\author[0000-0001-7126-4562]{T.F.L.L. Pinheiro}
\affiliation{Observat\'orio Nacional/MCTI, R. General Jos\'e Cristino 77, CEP 20921-400, RJ, Brazil}
\email[show]{franciscopinheiro@on.br}  

\begin{abstract}
The classical Roche limit is often used as a boundary between ring-dominated and satellite-dominated regions, although it is derived for a self-gravitating fluid satellite orbiting a spherical primary. This assumption is poorly suited to ring-bearing minor bodies, which are often triaxial. Here, we compute the equilibrium figures and disruption limits of inviscid fluid satellites orbiting triaxial primaries, modelling their gravity fields through $J_2$ and $C_{22}$. The primary's oblateness and equatorial asymmetry shifts the disruption limit outward relative to the classical Roche limit, with $C_{22}$ dominating the correction at sufficiently high values, which require highly oblate primaries. The disruption location also depends on the primary's orientation, defining a radial disruption region rather than a single limit in non-double-synchronous systems. Increasing orbital inclination shifts the limit inward, significantly so only at high inclinations. Symbolic regression provides an empirical expression for the disruption limit around triaxial primaries, which we apply to ring-bearing minor bodies. The correction is most relevant for Haumea and Chariklo, whereas the rings of Chiron and Quaoar located beyond the classical Roche limit remain outside the disruption region for realistic particle densities. This indicates that treating the disruption limit as a strict boundary between ring-dominated and satellite-dominated regions is an oversimplification, since the long-term survival of a ring also depends on collisional evolution, particle properties, and resonant confinement driven by the primary's non-axisymmetric gravity field.
\end{abstract}
\keywords{planets and satellites: dynamical evolution and stability, planets and satellites: rings, planets and satellites: formation}

\section{Introduction} \label{sec:intro}

The equilibrium figure of a perfect, self-gravitating, incompressible fluid satellite orbiting a spherical planet is a classical problem addressed by \citet{Roche1844} and \cite{Darwin1906}. In this idealised framework, the satellite assumes a triaxial ellipsoidal equilibrium figure under the combined effects of self-gravity, rotation, and planetary tides, with semi-axes $a\geq b\geq c$. The corresponding axis ratios, $\alpha=c/a$ and $\beta=b/a$, decrease as the satellite approaches the planet.

Thus, an inward-migrating fluid satellite should become progressively more elongated until it reaches a critical distance below which no equilibrium figure exists and tidal disruption occurs, producing a disk of material \citep[see, e.g.,][]{Madeira2023A}. This distance is the fluid Roche limit, written as \citep{Chandrasekhar1969}
\begin{equation}
a_R=2.445\left(\frac{\rho_p}{\rho}\right)^{1/3}R_p ,\label{eq_classical}
\end{equation}
where $\rho_p$ and $R_p$ are the bulk density and physical radius of the primary, and $\rho$ is the satellite bulk density.

Subsequent studies extend this problem by adopting more realistic assumptions for both the satellite and the primary. \citet{Darwin1906}, for example, considers a fluid Maclaurin spheroid as the primary, while \citet{Chandrasekhar1969,Kinoshita1972,Scheeres2006} investigate the equilibrium of two finite rigid bodies with comparable masses. Other works relax the fluid assumption for the satellite, treating it instead as a granular or cohesive body \citep{Dobrovolskis1990,Holsapple2006,Holsapple2008,Sharma2009}. In these cases, tidal disruption occurs at distances smaller than the fluid Roche limit, defining a rigid Roche limit. While \citet{Dobrovolskis1990,Holsapple2006,Holsapple2008} assume a spherical primary, \citet{Sharma2009} derives the equilibrium equations for a triaxial primary and shows that primary oblateness shifts the disruption limit\footnote{In this work, for clarity, we reserve the term Roche limit for the classical fluid Roche limit, derived for a perfect, self-gravitating, incompressible fluid satellite orbiting a spherical primary. For more realistic scenarios involving non-fluid satellites or non-spherical primaries, including those obtained in our simulations, we refer instead to the disruption limit.} outwards.

In general, the Roche limit has been treated in the literature as a sharp boundary between an inner ring-dominated region, where satellites are tidally disrupted and rings cannot efficiently coagulate, and an outer satellite-dominated region, where rings composed of macroscopic particles can accrete into moons. This interpretation is only approximate. Several planetary satellites lie inside the fluid Roche limit because they are not fluid bodies. A particularly striking example is provided by Saturn's moons Pan and Daphnis, which orbit within gaps in the A ring despite being located inside the Roche limit. Moreover, ring coagulation also depends on the ring mass distribution, impact velocities, optical depth, and particle size distribution \citep[e.g.][]{Karjalainen2007}. Thus, the survival of rings or satellites cannot be inferred from the classical Roche limit alone.

This limitation becomes even more relevant for minor-body systems, whose primaries often have highly non-spherical shapes. The discovery of the two massive rings Q1R and Q2R around Quaoar, both located well beyond the Roche limit for realistic particle densities \citep{Morgado2023,Pereira2023}, has raised questions about the applicability of the Roche limit concept to such systems. Additional mechanisms, beyond the classical Roche limit, may therefore control the stability of rings around minor bodies. In this context, \citet{Madeira2026} propose that Quaoar's rings remain stable as rings because Quaoar's equatorial asymmetry and the gravitational perturbations from Weywot sustain a high radial velocity dispersion among ring particles, inhibiting coagulation \citep[see also][]{Dinsmore2027}.

In this work, we use a numerical procedure based on \citet{Holsapple2006,Holsapple2008} to explore the disruption limit of an inviscid fluid satellite around a triaxial primary, defined as the innermost orbital distance at which an equilibrium figure of the satellite can exist. Our goal is to investigate how the triaxial shape of the primary modifies this limit relative to the classical Roche limit and to examine the implications of these modifications for the ring systems of Chariklo, Haumea, Chiron, and Quaoar. 
Section~\ref{sec_methods} presents our numerical method. Section~\ref{sec_results} examines the effects of the primary's shape on the satellite's equilibrium figure and disruption limit, while Section~\ref{sec_density} explores the dependence of the disruption limit on the primary and satellite densities. In Section~\ref{sec_expression}, we employ a machine-learning algorithm to derive an empirical expression for the disruption limit around triaxial bodies. Finally, Section~\ref{sec_implications} discusses the implications for ring-bearing minor bodies and presents our conclusions.

\section{Methods}\label{sec_methods}
\begin{figure}
    \fig{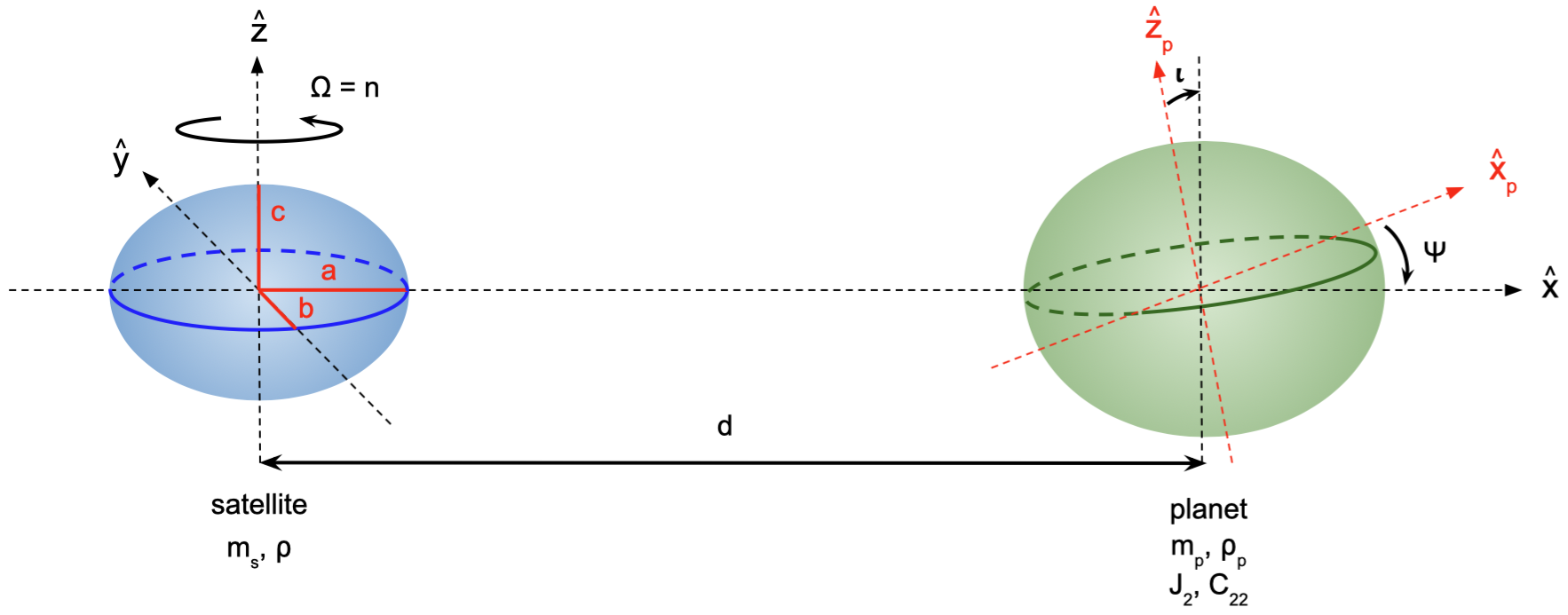}{1.0\textwidth}{}
  \caption{Schematic representation of the dynamical model, composed of a primary and a satellite with masses $m_p$ and $m_s$, and bulk densities $\rho_p$ and $\rho$, respectively. The centre-to-centre distance between the two bodies is $d$. The rotating frame $(x,y,z)$ is centred on the satellite and rotates about the satellite spin axis with angular frequency $\Omega$. The satellite is assumed to be synchronously rotating ($\Omega=n$), and its principal axes remain aligned with the rotating frame: the axes $x$, $y$, and $z$ coincide with the triaxial axes $a$, $b$, and $c$, respectively. The triaxial primary, whose gravitational field is described by the coefficients $J_2$ and $C_{22}$, is located along the $x$ axis. The primary-fixed frame $(x_p,y_p,z_p)$ is oriented such that $z_p$ coincides with the primary polar axis, while $x_p$ and $y_p$ lie in its equatorial plane. The angle $\iota$ defines the inclination of the primary polar axis with respect to the normal to the satellite orbital plane, whereas $\Psi$ defines the orientation of the primary's long axis within its equatorial plane relative to the satellite-primary geometry.}
  \label{fig:scheme}
\end{figure}

We consider a homogeneous inviscid fluid satellite of mass $m_s$ orbiting a triaxial primary of mass $m_p$ on a circular orbit with semi-major axis $d$ and mean motion $n=\sqrt{G(m_s+m_p)/d^3}$. For an inviscid fluid body, the equilibrium figure is a stationary shape in which self-gravity, rotation, and the tidal forces exerted by the primary are in hydrostatic balance \citep{Holsapple2006,Holsapple2008}. Here, we assume an ellipsoidal shape with semi-axes $a\geq b\geq c$, such that only specific combinations of these semi-axes satisfy the equilibrium condition at a given orbital distance. 

To describe the system, we adopt a satellite-centred rotating frame $(x,y,z)$, with the $x$-axis pointing along the satellite-primary direction and the $z$-axis normal to the orbital plane, and assume a spin-orbit locked configuration, $\Omega=n$ (Figure~\ref{fig:scheme}). The primary is also assumed to be triaxial, but its figure is taken as fixed and intrinsic, described by the gravitational coefficients $J_2$ and $C_{22}$. We define $\iota$ as the angle between the primary's polar axis, $\hat{z}_p$, and the normal to the satellite orbital plane, $\hat{z}$, which allows us to crudely account for an inclination of the satellite's orbit. The axes $(x_p,y_p,z_p)$ define the primary-fixed frame, with $x_p$ and $y_p$ lying in the equatorial plane of the primary. We further define $\Psi$ as the angle of the primary's long axis within this plane relative to the satellite-primary geometry. For $\iota=0^\circ$, the primary equatorial plane coincides with the satellite orbital plane, recovering the coplanar configuration.

Because an inviscid fluid cannot sustain shear stresses, its surface must coincide with an equipotential surface of the total effective potential \citep{Holsapple2006,Holsapple2008}. We rely on this condition to determine the equilibrium figures of the satellite and, consequently, to find the disruption location. For this reason, we must treat both $\Psi$ and $\iota$ as fixed parameters. The resulting shapes should therefore be interpreted as instantaneous equilibria, corresponding to the figures that a perfect inviscid-fluid satellite would assume if it responded instantaneously to the rotating orientation of the primary's gravitational field. In this sense, our approach corresponds to a static-tide approximation \citep{Ferraz2008}.

To numerically identify the instantaneous equilibrium figures, we use a surface-based equipotential fitting method that, for each distance $d$, identifies the ellipsoidal surface on which the total effective potential acting on the satellite is approximately constant. Thus, our method differs from that of \citet{Sharma2009}, which obtains equilibrium by requiring the stress state to be purely hydrostatic.

The total effective potential acting on the satellite is given by \citep{Chandrasekhar1969,Dobrovolskis1990}
\begin{equation}
U=\pi \rho G\left(-A_0+A_x x^2+A_y y^2+A_z z^2\right)
-\frac{\Omega^2}{2}(x^2+y^2)+U_{\rm primary}, \label{potential}
\end{equation}
where the first term is the satellite self-gravitational potential, the second is the centrifugal potential, and the last is the potential generated by the primary.

The coefficient $A_0$ is given by \citep{Chandrasekhar1969}
\begin{equation}
A_0=abc\int_0^{\infty} \frac{\mathrm{d}u}{
\sqrt{\left(a^2+u\right)\left(b^2+u\right)\left(c^2+u\right)}} ,
\end{equation}
whereas the coefficients $A_i$, with $i=x,y,z$, are \citep{Chandrasekhar1969}
\begin{equation}
A_i=abc \int_0^{\infty}\frac{\mathrm{d} u}{\left(a_i^2+u\right)
\sqrt{\left(a^2+u\right)\left(b^2+u\right)\left(c^2+u\right)}} .
\end{equation}

The primary potential is written as the sum of the monopole term and the contributions associated with oblateness ($J_2$) and equatorial ellipticity ($C_{22}$), denoted by $U_{\rm direct}$ \citep[e.g.,][]{Celletti2017,Madeira2025}, together with the indirect term $U_{\rm indirect}$. To account for an arbitrary inclination of the primary's polar axis relative to the orbital normal, we introduce primary-centred coordinates rotated so that the $z_p$-axis coincides with the primary's polar axis:
\begin{equation}
x_p = (d-x)\cos\iota-z\sin\iota,
\end{equation}
\begin{equation}
y_p = y,
\end{equation}
and
\begin{equation}
z_p = (d-x)\sin\iota+z\cos\iota.
\end{equation}
In this frame, the direct gravitational potential generated by the primary is
\begin{equation}
\begin{aligned}
U_{\rm direct} =
&-\frac{G m_p}{r_p}
-\frac{1}{2}G m_p J_2
\left(\frac{3m_p}{4\pi\rho_p}\right)^{2/3}
\frac{x_p^2+y_p^2-2z_p^2}{r_p^5}-3G m_p C_{22}
\left(\frac{3m_p}{4\pi\rho_p}\right)^{2/3}
\frac{\xi}{r_p^5}, \label{pot_geral}
\end{aligned}
\end{equation}
where $r_p=\sqrt{(d-x)^2+y^2+z^2}$ and
\begin{equation}
\xi=\left(x_p^2-y_p^2\right)\cos(2\Psi)+2x_py_p\sin(2\Psi).
\end{equation}
Because the satellite-centred frame is non-inertial, the indirect term is associated with the acceleration produced by the primary at the origin of the satellite-centred frame. In its general three-dimensional form, the indirect term takes the form \citep[e.g.,][]{Murray1999}:
\begin{equation}
U_{\rm indirect}=-\nabla U_{\rm direct}(0,0,0)\cdot\mathbf{r},
\end{equation}
where $\mathbf{r}=(x,y,z)$.

The equipotential surface is obtained numerically as follows. We first fix the primary properties ($m_p$, $\rho_p$, $J_2$, $C_{22}$, and $\Psi$), together with the mass and density of the satellite, from which we compute its equivalent radius $R_e$. Instead of treating the three semi-axes as independent quantities, we describe the satellite shape through the axis ratios $\alpha=c/a$ and $\beta=b/a$, fixing a value of $\alpha$. This allows us to relate the semi-axis $a$ to $\beta$ through the equation $R_e=a(\alpha\beta)^{1/3}$. Therefore, for each value of $\beta$, the semi-axis $a$ is recalculated, defining a different triaxial ellipsoid while keeping the satellite volume constant.

For each trial pair $(\beta,d)$, we construct the corresponding ellipsoidal surface and sample it with $N=20\times20$ points using the polar and azimuthal angles $(\theta,\phi)$ of the classical spherical-coordinate parametrization. Each sampled point has coordinates $(x_i,y_i,z_i)$, at which the total effective potential $U_i$ is evaluated per Equation~\ref{potential}. The mean potential over the sampled surface is computed as:
\begin{equation}
\langle U\rangle=\frac{1}{N}\sum_{i=1}^{N}U_i .
\end{equation}
We then adjust $\beta$ and $d$ by minimising the dispersion of the potential over the sampled surface:
\begin{equation}
\min_{\beta,d}
\sum_{i=1}^{N}
\left[
U(x_i,y_i,z_i;\alpha,\beta,d)-\langle U\rangle
\right]^2 . \label{eq_minim}
\end{equation}

For each primary-satellite pair, we explore 50 values of $\alpha$, uniformly distributed over the interval $0.02\leq\alpha\leq0.98$, obtaining the corresponding equilibrium values of $\beta$ and $d$ for each value of $\alpha$. These solutions define a curve in the $(\alpha,d)$ plane (Figure~\ref{fig:roche}a,c). The disruption limit is then determined by selecting the smallest value of $d$ among the 50 equilibrium solutions obtained, which corresponds to the minimum of the curve in the $(\alpha,d)$ plane. At smaller distances, no combination of $\alpha$ and $\beta$ yields an ellipsoidal equipotential surface.

\section{Dependence on primary shape and orientation}\label{sec_results}
\begin{figure*}
  \gridline{
    \fig{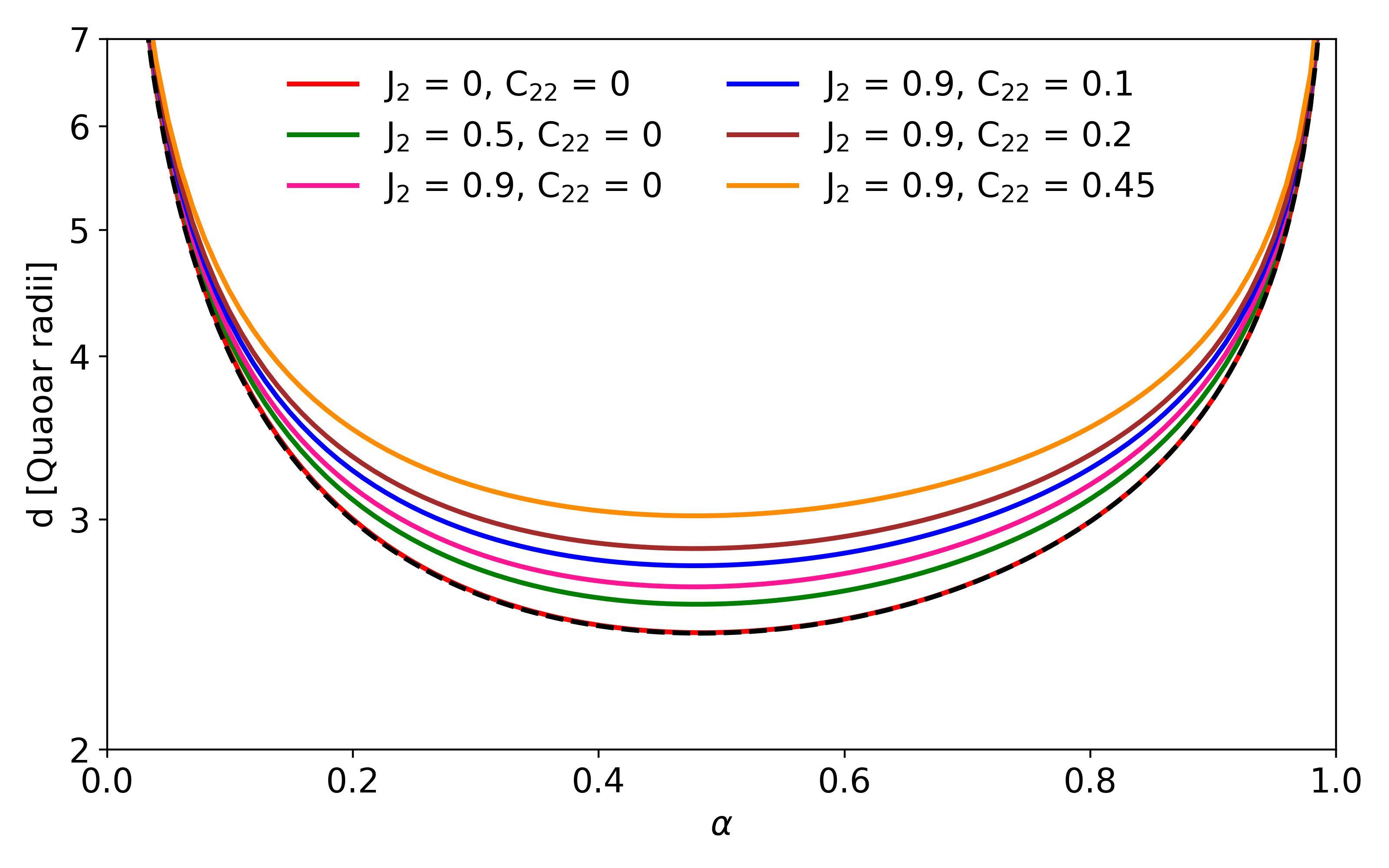}{0.48\textwidth}{(a)}
    \fig{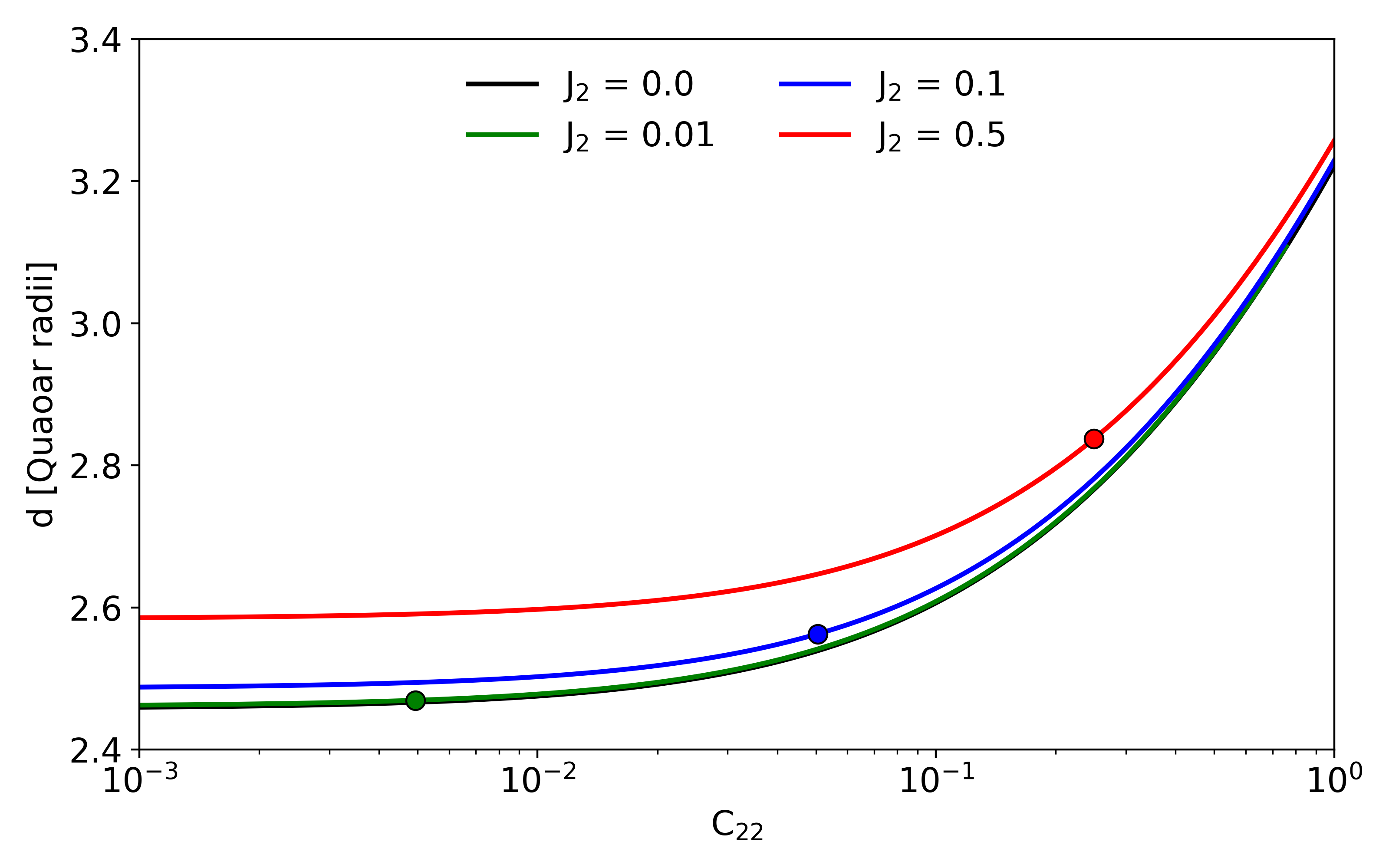}{0.48\textwidth}{(b)}
  }
\gridline{
    \fig{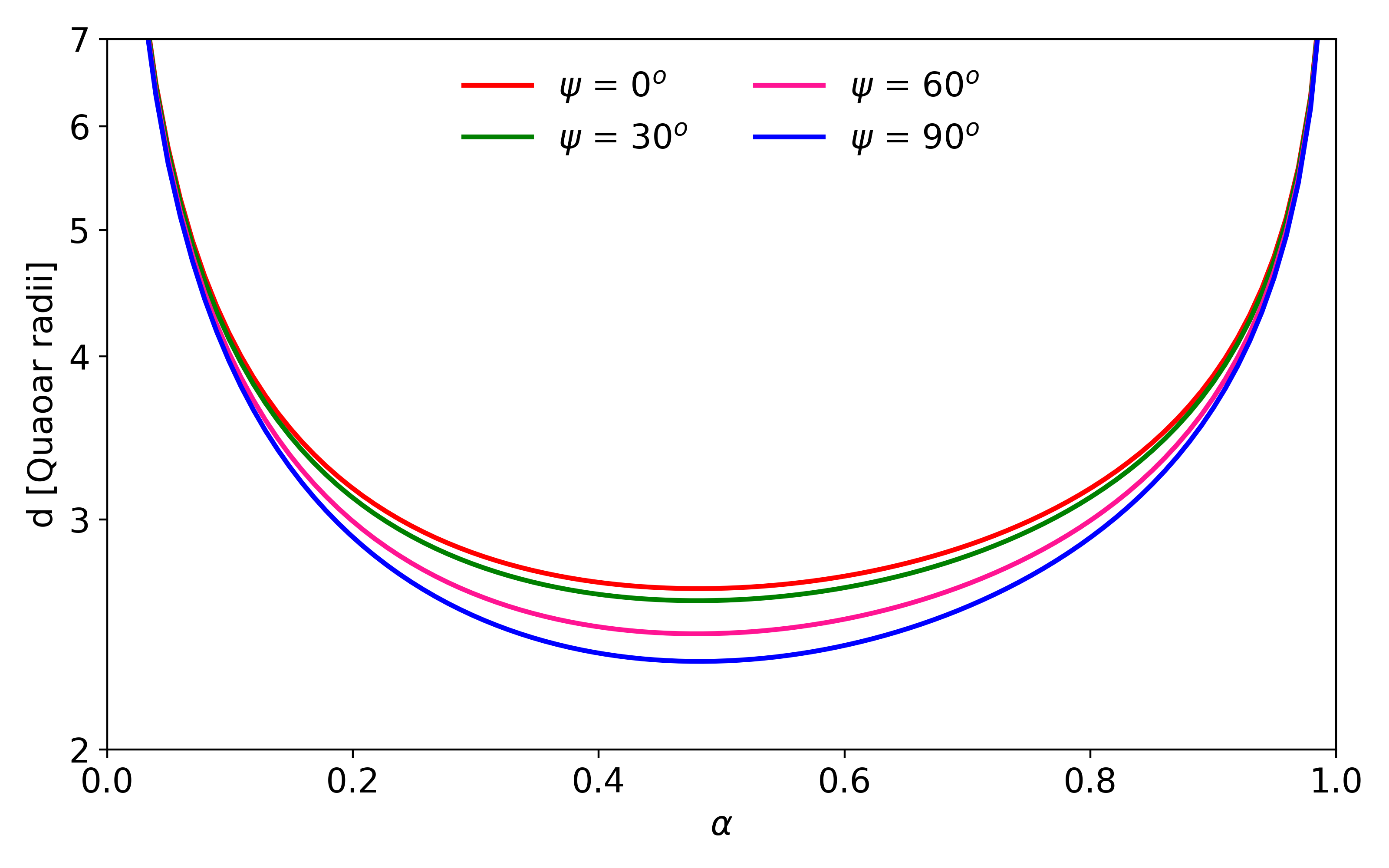}{0.48\textwidth}{(c)}
    \fig{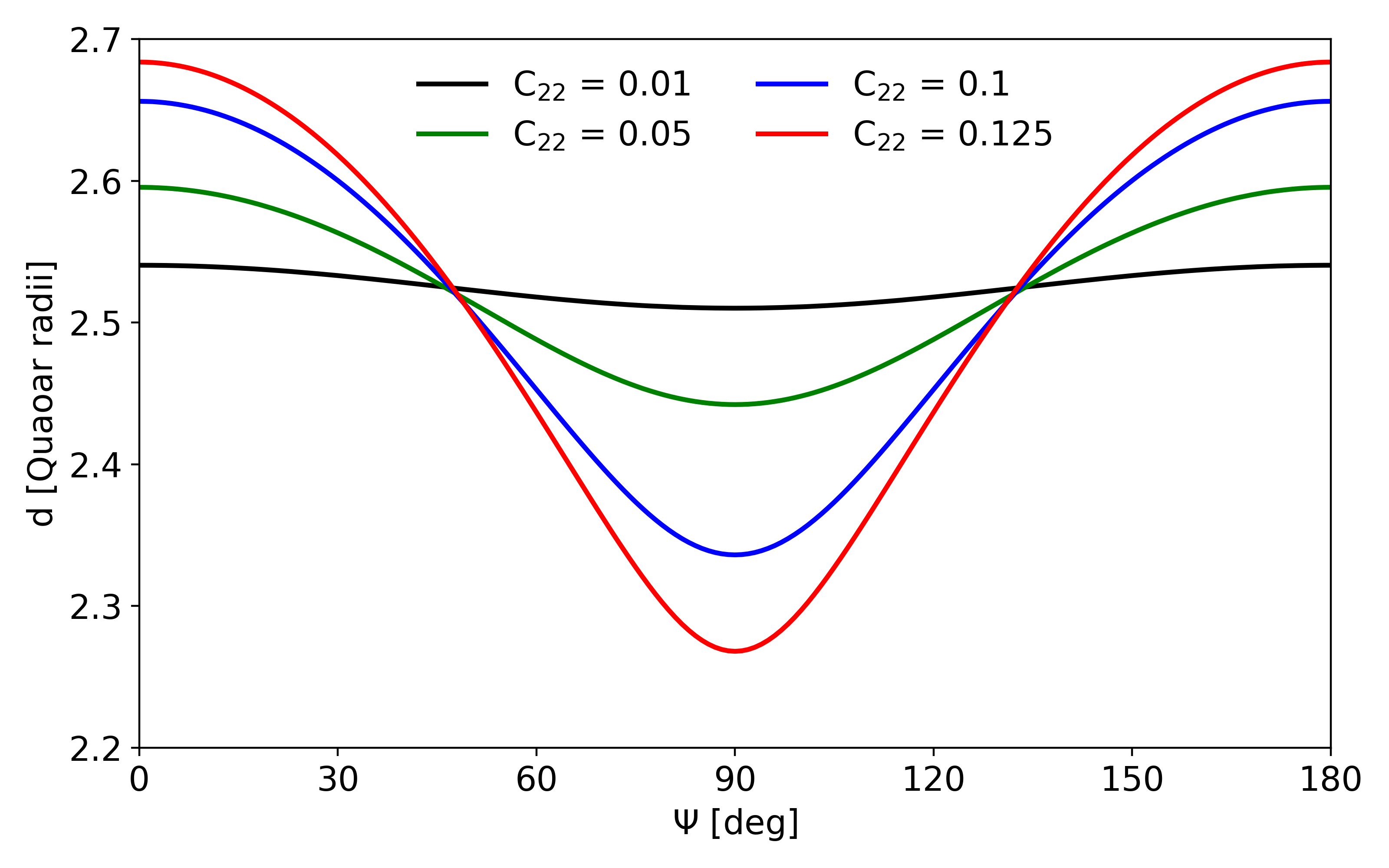}{0.48\textwidth}{(d)}
  }
\gridline{
    \fig{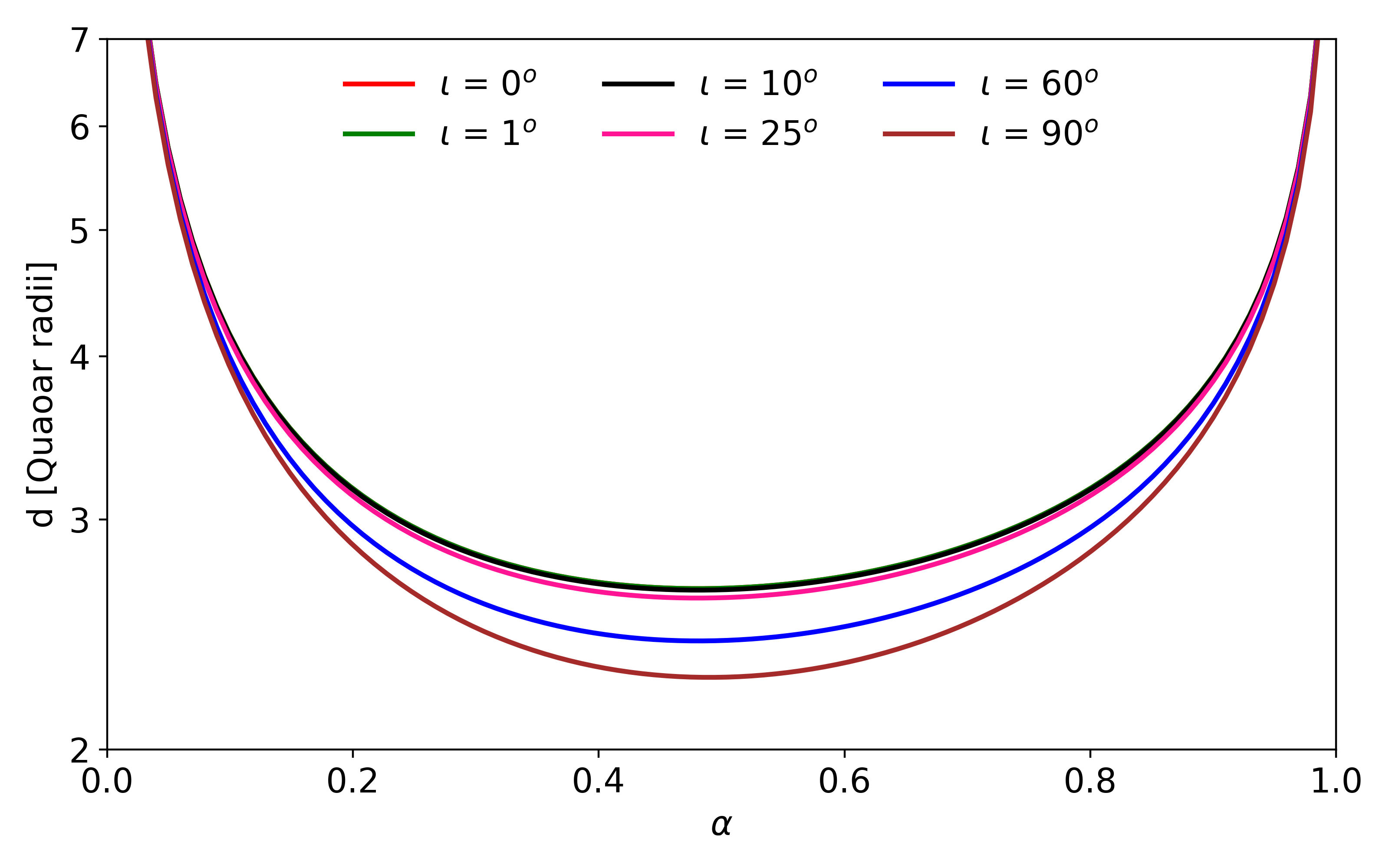}{0.48\textwidth}{(e)}
    \fig{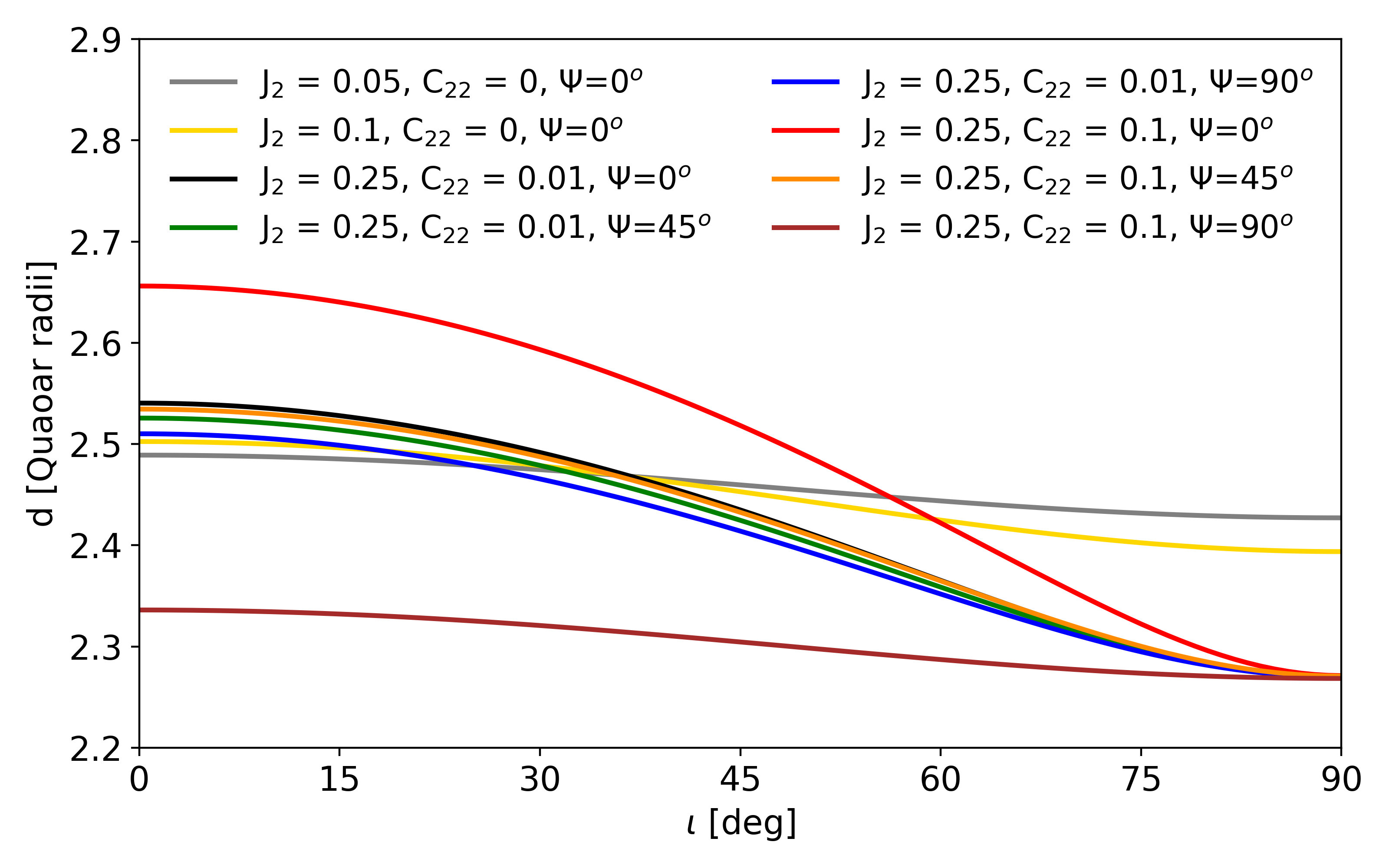}{0.48\textwidth}{(f)}
  }
  \caption{Disruption limit and satellite equilibrium figures as a function of the primary's quadrupolar gravity field and orientation. Panels (a,b) show the effect of $J_2$ and $C_{22}$ for $\Psi=\iota=0^\circ$: (a) equilibrium distance $d/R_p$ as a function of $\alpha=c/a$, and (b) disruption limit as a function of $C_{22}$ for different $J_2$. Panels (c,d) show the effect of $\Psi$: (c) equilibrium curves for $J_2=0.25$, $C_{22}=0.1$, and $\iota=0^\circ$, and (d) disruption limit for different $C_{22}$, with $J_2=0.25$ and $\iota=0^\circ$. Panels (e,f) similarly show the effect of $\iota$: (e) equilibrium curves for $J_2=0.25$, $C_{22}=0.1$, and $\Psi=0^\circ$, and (f) disruption limits for different combinations of $J_2$, $C_{22}$, and $\Psi$. The colored dots in panel (b) indicate the maximum values of $C_{22}$ allowed for a triaxial primary, $C_{22}\leq0.5J_2$. Following the physical properties of Quaoar, we assume $m_p=1.2\times10^{21}$~kg, a satellite-to-primary mass ratio of $m_s/m_p=10^{-3}$, and equal bulk densities for the primary and satellite, $\rho_p=\rho=1676~{\rm kg\,m^{-3}}$ \citep{Proudfoot2025}.}
  \label{fig:roche}
\end{figure*}

We start our analysis by examining the effect of primary oblateness ($J_2$) and equatorial asymmetry ($C_{22}$) for fixed orientation angles $\Psi=\iota=0^\circ$ (Figure~\ref{fig:roche}a,b), and then vary $\Psi$ (Figure~\ref{fig:roche}c,d) and $\iota$ (Figure~\ref{fig:roche}e,f) to evaluate the effects of the primary bulge orientation and inclination, respectively. Panels (a), (c), and (e) show the normalized equilibrium distance $d/R_p$ as a function of the satellite axis ratio $\alpha$. For each fixed value of $\alpha$, we numerically solve Equation~\ref{eq_minim}, varying $\beta$ and $d$ to minimize the dispersion of the total effective potential over the satellite surface. Each point along the curves therefore represents an equilibrium figure obtained through the procedure described in Section~\ref{sec_methods}. The minimum distance along each curve defines the disruption limit, which is shown in panels (b), (d), and (f) as a function of $C_{22}$, $\Psi$, and $\iota$, respectively. Our numerical solution for $J_2=C_{22}=0$ exactly reproduces the classical spherical-primary curve \citep[dashed line;][]{Holsapple2006}, validating the method. 

For fixed $\Psi=\iota=0^\circ$, the satellite-shape curves (Figure~\ref{fig:roche}a) closely resemble the curve obtained for a spherical primary. In particular, for a given distance, two equilibrium solutions are found, each belonging to a different family, one stable and the other unstable\footnote{Here, stability refers to the response of the equilibrium figure to small perturbations \citep{Chandrasekhar1969}. A satellite belonging to the stable family can readjust its shape to a new equilibrium configuration, whereas a satellite belonging to the unstable family is driven toward disruption.}. These two families converge at $\alpha_{\min}$ and $\beta_{\min}$, which define the satellite shape at the minimum equilibrium distance, i.e., at the disruption limit. The unstable family is highly flattened ($\alpha<\alpha_{\min}$) and becomes more spherical ($\alpha$ increases) as the satellite approaches the primary, while the stable family ($\alpha>\alpha_{\min}$) is initially more spherical and becomes progressively more elongated at smaller distances ($\alpha$ decreases). The latter corresponds to the Darwin ellipsoids and is the family of solutions that a fluid satellite is expected to assume \citep{Chandrasekhar1969,Madeira2026EPJST}. The satellite shape at the disruption limit is also unchanged with respect to the spherical-primary case, with $\alpha_{\min}=0.48$ and $\beta_{\min}=0.51$.

The main effect of non-zero $J_2$ and $C_{22}$ for $\Psi=\iota=0^\circ$ is to shift the satellite-shape curve outwards. This occurs because increasing $J_2$ and $C_{22}$ strengthens the in-plane tidal forcing produced by the primary. Therefore, the more oblate and equatorially asymmetric the primary is, the farther from the primary the satellite reaches the critical configuration for tidal disruption (Figure~\ref{fig:roche}b). A similar trend is found by \citet{Sharma2009}, who shows that the equilibrium distance of a satellite with given $\alpha$ and $\beta$ increases with the primary's oblateness and equatorial ellipticity. 

To assess the relative effects of the gravitational coefficients, we first consider oblateness ($J_2$) alone and then evaluate the additional effect of equatorial asymmetry ($C_{22}$). For example, a highly oblate primary with $J_2=0.9$ and $C_{22}=0$ shifts the disruption limit from $2.445$ Quaoar radii in the spherical case to $2.765$ Quaoar radii (a 13\% increase). When $C_{22}$ is increased to $0.45$ while keeping $J_2=0.9$, the disruption limit shifts further outward to $3.020$ Quaoar radii (an additional 8\% increase). Although $J_2$ produces a larger change when the coefficients are varied independently, the contribution of $C_{22}$ becomes increasingly relevant for bodies with larger values of $C_{22}$ because it introduces in-plane shear components into the tidal field. Consequently, the satellite is not only stretched or compressed, as in the pure-$J_2$ case, but also twisted, making the equilibrium figure less stable at smaller distances from the primary.

As shown in Figure~\ref{fig:roche}b, the effect of $C_{22}$ on the disruption limit is minor for $C_{22}\lesssim10^{-3}$, but becomes increasingly important, and eventually dominant, for $C_{22}\gtrsim10^{-2}$. We must point out, however, that $C_{22}$ is not independent of $J_2$, since the triaxial figure of the primary imposes $C_{22}\leq0.5J_2$. Thus, a significant contribution from $C_{22}$ necessarily requires an oblate primary. For Saturn, the most oblate planet in the Solar System, its $J_2=0.016$ \citep{Iess2019} imposes a maximum $C_{22}\sim10^{-3}$, so that equatorial asymmetry should not have any significant effect on the disruption location of satellites around Solar System planets.

At smaller scales, however, minor planets and large asteroids can have shapes further from a spherical approximation than planets, while still being reasonably described as triaxial figures. Vesta, for example, has $J_2\sim0.07$, allowing values of $C_{22}$ within the range in which equatorial asymmetry could become relevant, although its actual value is only $C_{22}\sim0.003$ \citep{Park2025}. Haumea represents a more extreme case, with $J_2\sim0.26$ and $C_{22}\sim0.05$ \citep{Ortiz2017,Proudfoot2024,Staelen2026}, for which the effect of equatorial asymmetry on the disruption limit becomes noticeable, as discussed in Section~\ref{sec_smallbodies}.

The shear tidal torque introduced by $C_{22}$ is also expected to depend on the orientation of the primary's bulge, represented by the angles $\Psi$ and $\iota$. For both parameters, as for $J_2$ and $C_{22}$, varying their values shifts the satellite-shape curve in distance (Figure~\ref{fig:roche}c,e), while preserving its overall shape and the value of $\alpha_{\min}$ at which the two families of equilibrium solutions converge. Focusing first on $\Psi$, an important trend observed in Figure~\ref{fig:roche}d, reflecting the symmetry of the primary's ellipsoidal figure, is that the disruption location varies periodically with this angle, with a period of $180^\circ$.

As $\Psi$ increases from $0^\circ$ to $90^\circ$, the entire curve of equilibrium shapes (Fig.~\ref{fig:roche}c) moves inward, reaching its smallest equilibrium distances at $\Psi=90^\circ$, and then moves outward as $\Psi$ increases further. The smallest disruption limit is naturally found at $\Psi=90^\circ$, where the satellite's major equatorial axis is aligned with the primary's minor equatorial axis, while the largest values occur at $\Psi=0^\circ$ and $\Psi=180^\circ$, where the major axes of both bodies are aligned.

The amplitude of the variation in the disruption radius with $\Psi$ increases with $C_{22}$ because a larger $C_{22}$ strengthens the equatorial asymmetry of the primary's gravitational field, enhancing the dependence of the tidal interaction on the relative orientation of the primary and satellite. For $C_{22}=0.01$, for example, the disruption distance changes only slightly with $\Psi$, whereas for $C_{22}=0.125$, it decreases from approximately $2.69$ Quaoar radii at $\Psi=0^\circ$ to $2.27$ Quaoar radii at $\Psi=90^\circ$, a variation of about $0.42$ Quaoar radii.

At this point, it is worth emphasizing again that $\Psi$ is kept fixed owing to a limitation of our method. This corresponds to the special case of mutual tidal locking, as observed, for example, between Pluto and Charon \citep{Tholen1994}. In most cases, however, the primary figure rotates with respect to the secondary, so that both the satellite equilibrium figure and the disruption limit should, strictly speaking, vary with the primary's rotation. An instantaneous reshaping of the satellite with the primary's rotation would require the satellite to respond instantaneously to the external gravitational field. However, even within the idealised approximation of a fluid satellite adopted here, the response of the satellite to the external gravitational field should present a finite time delay due to internal dissipation \citep[e.g.][]{Goldreich1966,Hut1981,Rambaux2013,Correia2014}, inducing tidal stresses and internal heating instead of a truly instantaneous change of the satellite figure.

In real bodies, tidal cycles driven by primary's rotation may induce material fatigue and enhance pre-existing faults \citep{Hammond2018}, generate stripe fractures \citep{Rhoden2020}, or even produce partial melting in strongly forced bodies \citep{Tobie2025}. Therefore, the maximum disruption limit, obtained for $\Psi=0^\circ$, should be interpreted more cautiously as the outer boundary of a radial region where tidal stresses are expected to act most strongly on the satellite. The actual disruption location is then likely to lie within the radial interval between this maximum distance and the minimum disruption limit obtained at $\Psi=90^\circ$.

For the inclination parameter, the equilibrium-shape curve (Fig.~\ref{fig:roche}e) is also observed to move inward as $\iota$ increases from $0^{\circ}$ to $90^{\circ}$, indicating that inclined satellites should be disrupted closer to the primary than equatorial ones. This decrease in the disruption location, however, becomes relevant only for relatively high inclinations, $\iota\gtrsim20^{\circ}$, well beyond the inclinations of regular satellites in the Solar System. As shown in Figure~\ref{fig:roche}f, the rate at which the disruption location decreases with $\iota$ increases with $J_2$ and $C_{22}$. For instance, for a oblate primary with $J_2=0.05$, the disruption location varies from 2.48~Quaoar radii at $\iota=0^\circ$ to 2.43~Quaoar radii at $\iota=90^\circ$, whereas for a highly oblate primary with $J_2=0.25$, it varies from 2.54 to 2.27~Quaoar radii.

A particularly interesting feature is that different curves with the same $J_2=0.25$ tend to converge as $\iota$ approaches $90^\circ$. Although the disruption locations do not converge to exactly the same value for different values of $C_{22}$ and $\Psi$, the contribution of $C_{22}$ becomes progressively less relevant as the satellite approaches the polar configuration. For example, for $J_2=0.25$ and $\Psi=0^\circ$, the disruption location is approximately $2.270$ Quaoar radii for $C_{22}=0.01$ and $2.271$ Quaoar radii for $C_{22}=0.1$, despite the order-of-magnitude difference in $C_{22}$. This occurs because increasing $\iota$ changes which components of the primary's gravity field dominate the in-plane tidal field experienced by the satellite. In the polar configuration, the relevant in-plane tidal field is primarily controlled by the difference between the polar and equatorial axes ($c$ and $a$), rather than by the difference between the two equatorial axes ($a$ and $b$), as in the equatorial configuration. Consequently, the contribution associated with $J_2$ becomes dominant over that of $C_{22}$.

Taken together, the results of Figure~\ref{fig:roche} demonstrate that the disruption limit around triaxial primaries cannot be accurately described by a simple density scaling that does not account for the primary's shape and orientation. For the Solar System planets, their relatively low oblateness ($J_2\lesssim10^{-2}$) imposes low values of $C_{22}$, such that the correction due to equatorial asymmetry is minor and the classical picture is recovered. For minor bodies, however, the disruption location can be significantly modified. In such cases, the orientation-dependent effect of the primary on the satellite shape creates a radial region over which an inward-migrating fluid satellite may disrupt. The satellite inclination is also observed to affect this location, although relatively high inclinations are required for this effect to become significant. In Section~\ref{sec_expression}, we explore how these distances are modified with respect to the classical prediction for triaxial primaries. Next, we explore the limits of the density-scaling rule itself.

\section{Dependence on bulk densities}\label{sec_density}

\begin{figure*}
  \gridline{
    \fig{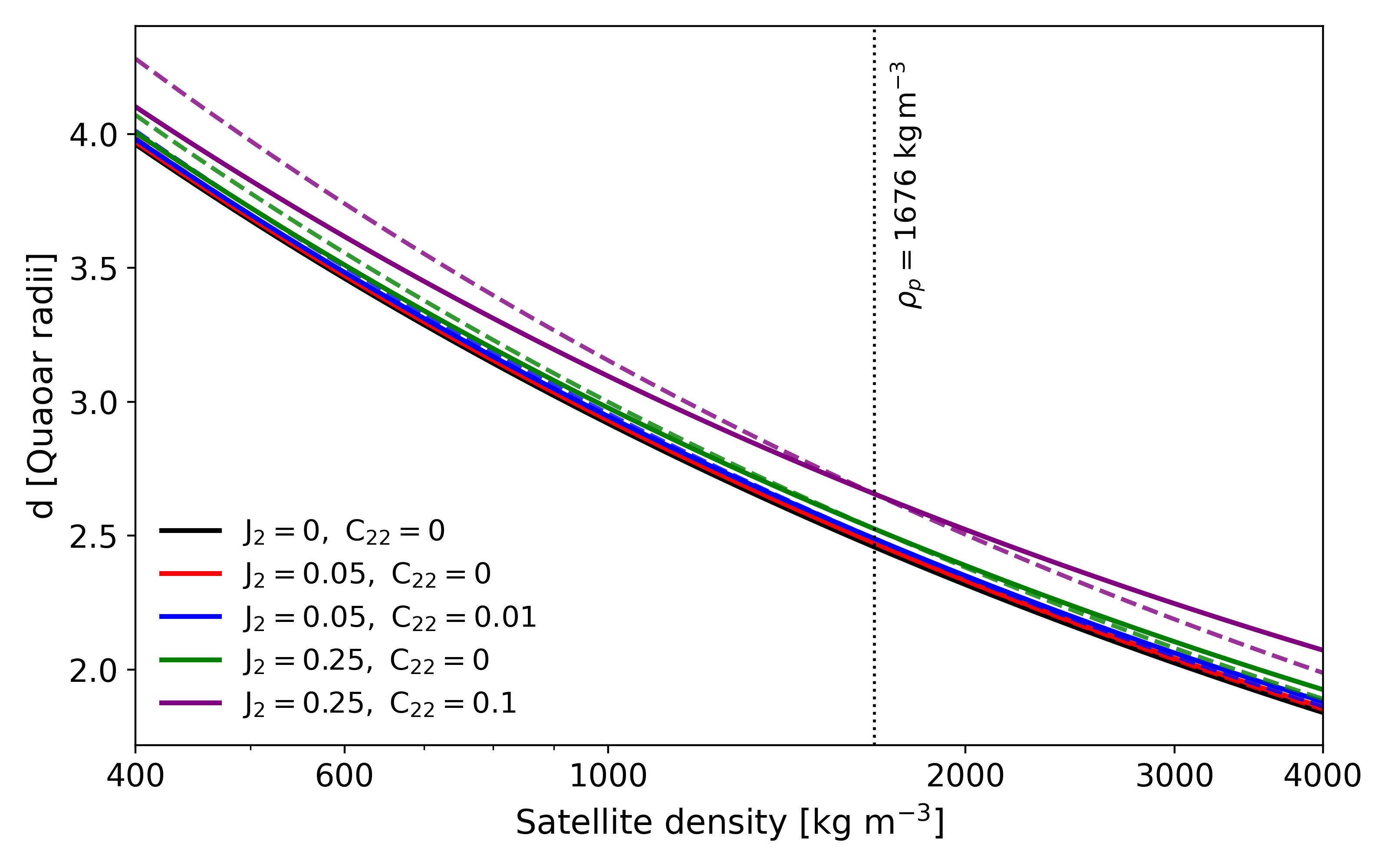}{0.48\textwidth}{(a)}
    \fig{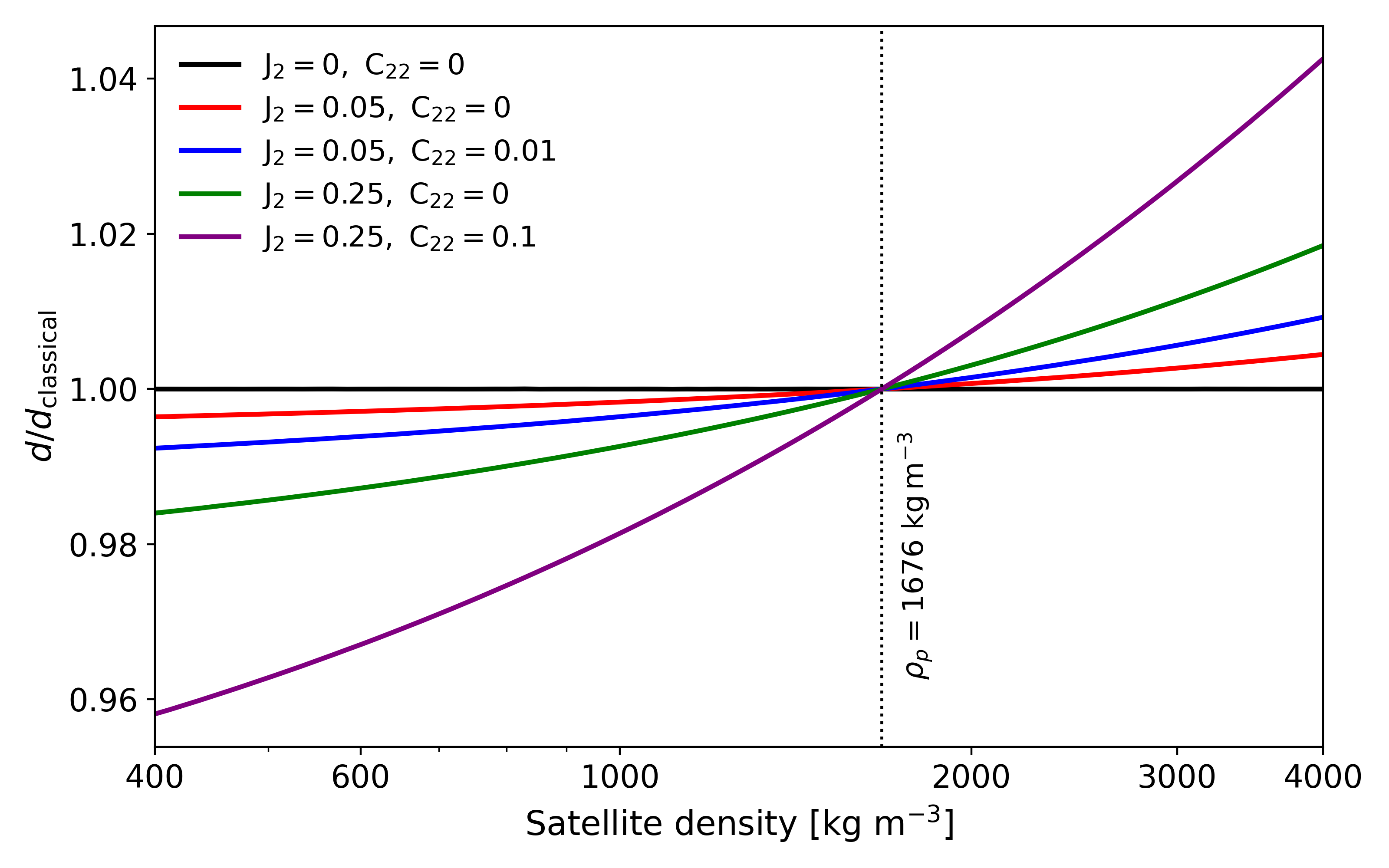}{0.48\textwidth}{(b)}
  }
\gridline{
    \fig{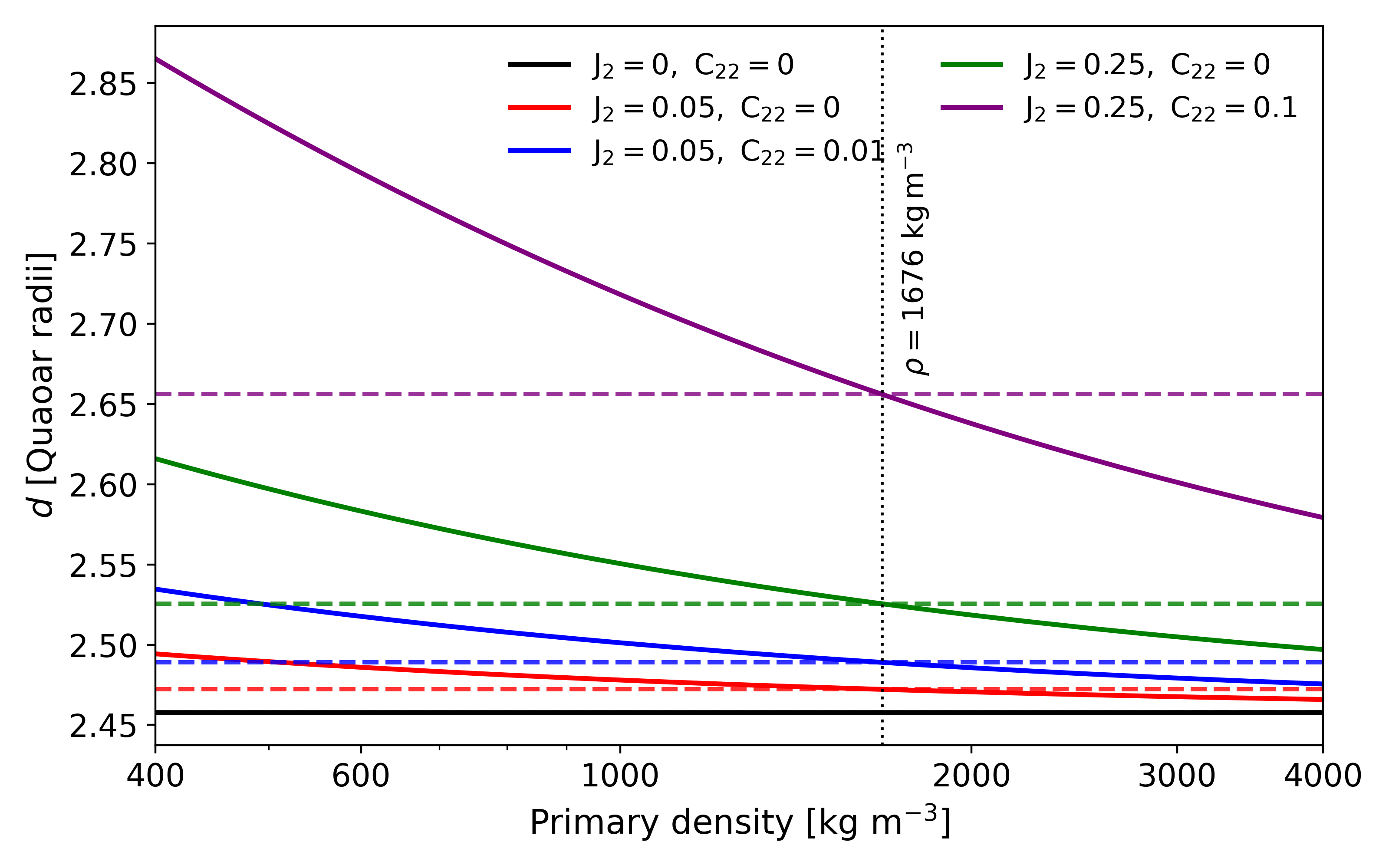}{0.48\textwidth}{(c)}
    \fig{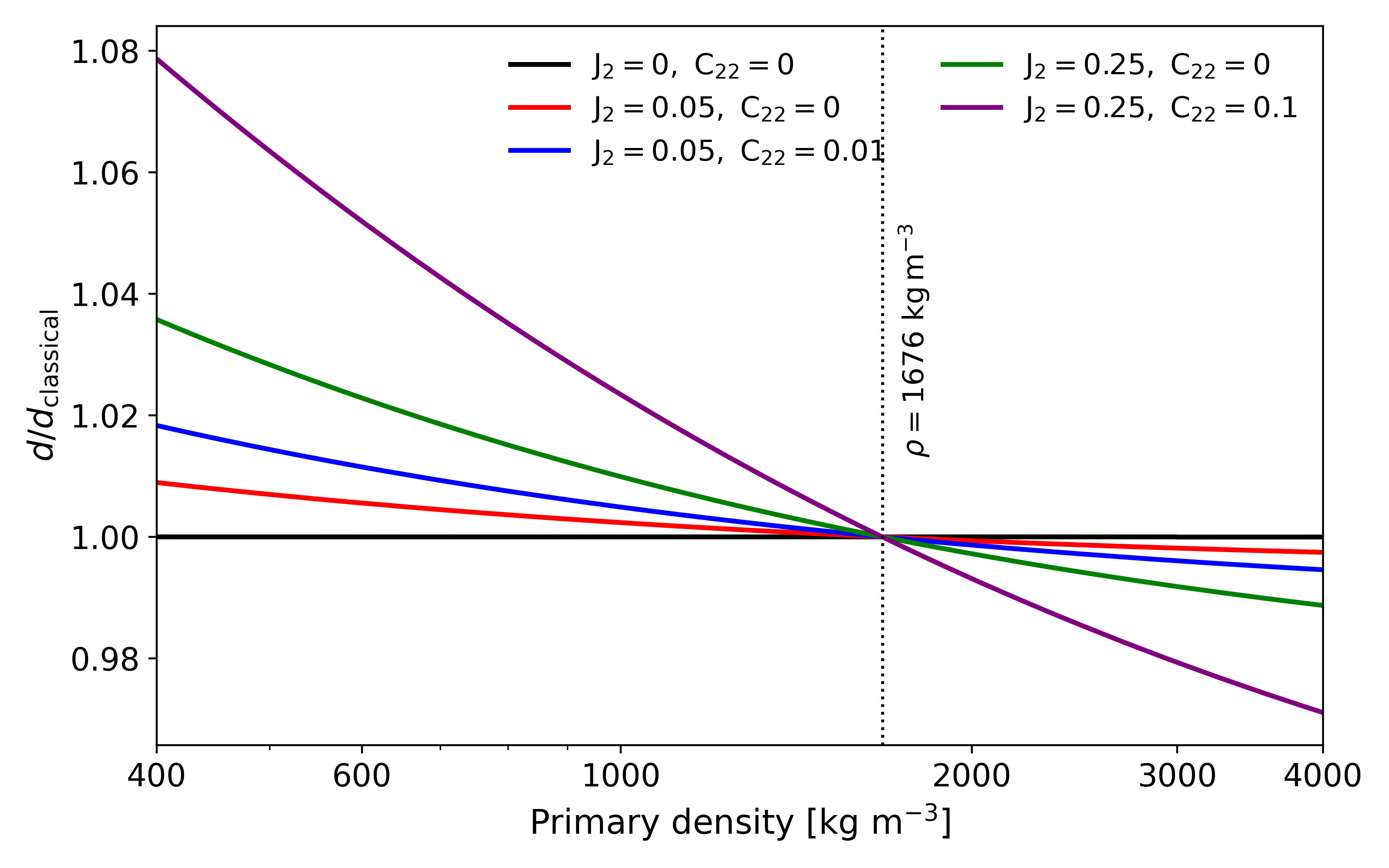}{0.48\textwidth}{(d)}
  }
  \caption{Dependence of the disruption limit on the satellite and primary densities. Panel (a) shows the disruption distance as a function of the satellite density, fixing the primary density at $\rho_p=1676~{\rm kg\,m^{-3}}$, while panel (b) shows the same results normalized by the classical disruption limit, $d_{\rm classical}$. Panel (c) shows the disruption distance as a function of the primary density, fixing the satellite density at $\rho=1676~{\rm kg\,m^{-3}}$, while panel (d) shows the same results normalized by $d_{\rm classical}$. The dashed lines mark the classical disruption limit, $d_{\rm classical}$. We assume $m_p=1.2\times10^{21}$~kg, a satellite-to-primary mass ratio of $m_s/m_p=10^{-3}$, and $\Psi=\iota=0^{\circ}$.}
  \label{fig:density}
\end{figure*}

To evaluate whether the density-ratio exponent of $1/3$ in the Roche limit prescription (Equation~\ref{eq_classical}) still holds when considering a triaxial primary, we performed experiments in which the primary and satellite masses were fixed while their bulk densities were varied. The results are shown in Figure~\ref{fig:density}. In panels (a) and (b), we explore different values of satellite density, while in panels (c) and (d), we explore different values of the primary density. To compare our results with the classical density-ratio scaling, we define the quantity $d_{\rm classical}$ (dashed lines in Figures~\ref{fig:density}a,c), given by
\begin{equation}
d_{\rm classical}=A\left(\frac{\rho_p}{\rho}\right)^{1/3}R_p, \label{eq_aclas}
\end{equation}
where $R_p$ is recalculated when the primary density is varied, and $A$ is set to the disruption location obtained numerically when $\rho=\rho_p$.

While the Roche limit prescription is recovered for a spherical primary ($J_2=C_{22}=0$), we find that the disruption radius departs from that predicted by the classical density-ratio scaling when the primary shape is taken into account, for variations in either the satellite or primary density. By construction, all curves satisfy $d/d_{\rm classical}=1$ at $\rho=\rho_p$, with deviations increasing away from this reference value. These results indicate that the effects of the primary's shape cannot be captured by simply modifying the coefficient $A$ in the classical Roche limit, but require an additional correction with a distinct dependence on the density ratio.

For variations in the satellite density, we obtain disruption radii interior to those predicted by Equation~\ref{eq_aclas} for $\rho<\rho_p$, and exterior to them for $\rho>\rho_p$. For variations in the primary density, in turn, $d_{\rm classical}$ expressed in Quaoar radii remains constant with $\rho_p$, since the primary radius is recalculated according to $R_p\propto\rho_p^{-1/3}$ for a fixed primary mass. In contrast, the numerically determined disruption radius decreases with increasing $\rho_p$. Consequently, the classical scaling overestimates the disruption distance for low-density satellites and high-density primaries, while underestimating it for high-density satellites and low-density primaries.

This mismatch with the classical density-ratio scaling increases with $J_2$ and $C_{22}$, indicating an explicit dependence of the additional correction on the primary's quadrupolar gravity field. A comparison between the isolated variations in the satellite and primary densities shows that the deviations are larger when the primary density is varied. For a primary with $J_2=0.25$ and $C_{22}=0.1$, for example, the deviations from $d_{\rm classical}$ reach $\sim4\%$ for the extreme satellite densities considered ($400~{\rm kg\,m^{-3}}$ and $4000~{\rm kg\,m^{-3}}$), whereas for variations in the primary density, the deviation reaches $\sim8\%$ at $\rho_p=400~{\rm kg\,m^{-3}}$.

The larger deviations for the primary density arise because varying it also changes the primary radius, which directly affects the quadrupolar terms of the primary gravitational potential (Equation~\ref{pot_geral}). In contrast, the effect of the satellite density is more indirect, since it does not modify the tidal field generated by the primary. Instead, changing the satellite density changes the distance at which its self-gravity balances the primary tidal field and rotation, thereby changing the relative contribution of the non-spherical terms to this balance.

In summary, we find that the classical density scaling is a particular result of the spherical-primary approximation rather than a universal scaling of the disruption limit, becoming increasingly inaccurate as $J_2$ and $C_{22}$ increase. These results highlight the need for a more general prescription that accounts for both the density ratio and the primary's non-spherical gravity field. In the following section, we use symbolic regression to derive an empirical expression for the disruption limit that incorporates these effects.

\section{An empirical expression for the disruption limit}\label{sec_expression}

A symbolic regression approach was employed to derive a disruption-limit expression that corrects the classical Roche limit for the effects of $C_{22}$, $J_2$, and $\Psi$. Here, we neglect the effect of $\iota$, as it becomes significant only at high orbital inclinations. To construct the symbolic regression model, we performed 10,000 numerical simulations for each of four hypothetical systems  resembling Chariklo, Quaoar, Haumea, and Chiron. Equal densities were assumed for the primary and satellite bodies and the physical parameters 
for both bodies were selected within the ranges reported in Table~\ref{Tab_equations}.

The coefficients $J_2$ and $C_{22}$ were computed by randomly varying the semi-axes of the primary's triaxial figure, $a_p$, $b_p$, and $c_p$, while keeping the equivalent radius ($R_p=\sqrt[3]{a_pb_pc_p}$) constant, thereby maintaining a constant density \citep{Balmino1994}:
\begin{equation}
    J_2 = \frac{a_p^2 + b_p^2 - 2c_p^2}{10R_p^2}
    \qquad
    \text{and}
    \qquad
    C_{22} = \frac{a_p^2 - b_p^2}{20R_p^2}.
\label{Eq_J2_C22}
\end{equation}
To ensure physically plausible primary shapes, the ellipsoid axis ratios were restricted to $a_p/c_p<4$ and $a_p/b_p<4$, with $a_p\geq b_p\geq c_p$. This upper limit is approximately twice the estimated $a_p/c_p$ ratio of Haumea \citep{Ortiz2017,Proudfoot2024,Staelen2026}. Given the symmetry of the disruption radius with respect to $\Psi$ (Figure~\ref{fig:roche}d), $\Psi$ was randomly sampled over $0^\circ\leq\Psi\leq90^\circ$.

Additionally, we generated a new dataset comprising 10,000 hypothetical systems with varying sizes, masses, and densities for both the primary and satellite bodies. For each system, the primary's density was randomly sampled from the range $\rho_p = 1000-4000~{\rm kg\,m^{-3}}$, while its equivalent radius was fixed at $R_p=500$~km. The primary's mass was then calculated from its density and equivalent radius, and the coefficients $J_2$ and $C_{22}$ were determined from the dimensions of the ellipsoid, whose semi-axes were allowed to vary between $200$ and $1000$~km. For the satellite, its density was also randomly sampled from the range $\rho = 1000-4000~{\rm kg\,m^{-3}}$, its mass was assumed to be $10^{-3}$ times that of the primary, and its radius was subsequently calculated from its mass and density. This satellite-to-primary mass ratio is motivated by the relatively large satellites found around some minor bodies, such as Haumea and Quaoar, whose satellites have mass ratios of the order of $10^{-3}$ \citep{Ragozzine2009,Margoti2026}.

Symbolic regression is a machine learning technique that searches for an analytical expression capable of accurately describing a given dataset \citep[see e.g.][]{Lemos2023,Pinheiro2026}. Candidate equations are represented as tree structures, in which variables, constants, and mathematical operators are organized as interconnected nodes and branches. To identify the best-fitting expression, the algorithm initializes a population of candidate equations and iteratively evolves them through genetic operations, including mutation, selection and crossover \citep{Augusto2000}. 

The dataset was randomly divided into training and testing subsets, with 80\% of the samples used for training and the remaining 20\% for model evaluation. The symbolic regression model was implemented using the PySR package \citep{Cranmer2023}, employing 50 independent populations evolved over 1,000 iterations. To control model complexity, each candidate expression was limited to a maximum tree size of 30 nodes. 

For the first four datasets, representing systems analogous to Chariklo, Quaoar, Haumea, and Chiron, a separate disruption-limit equation was fitted to each dataset using $J_2$, $C_{22}$, $\cos(\Psi)$, and $\sin(\Psi)$ as input features. For the fifth dataset, which included systems with different bulk densities for the primary and satellite bodies, the density ratio $\rho_p/\rho$ was added to the feature space. In all cases, the symbolic regression models were allowed to combine the input features using addition, subtraction, multiplication, and exponentiation.

\begin{table*}[h!]
\centering
\small
\setlength{\tabcolsep}{3.5pt} 
    \begin{tabular}{lccccclcc}
        \hline
            Body & Mass (kg) & ${\rm a_p,b_p,c_p}$ (km) & $R_{p}$ (km) & $T$ (h) &
            Disruption equation & RMSE & $R^2$ \\
            \hline
            Chiron
            & $4.8\times10^{18}$
            & $50$--$200$
            & $98$
            & $5.918$
            & $2.463 + 3.502C_{22}(\cos\Psi-0.677) + 0.261J_2$
            & $0.0063$
            & $0.99$ \\
            
            Chariklo
            & $8.0\times10^{18}$
            & $50$--$200$
            & $126$
            & $7.004$
            & $2.461 + 3.118C_{22}(\cos^2\Psi-0.518) + 0.256J_2$
            & $0.0037$
            & $0.99$ \\
            
            Haumea
            & $4.0\times10^{21}$
            & $400$--$1600$
            & $797.5$
            & $3.915$
            & $2.463 + 3.500C_{22}(\cos\Psi-0.674) + 0.259J_2$
            & $0.0094$
            & $0.99$ \\
            
            Quaoar
            & $1.2\times10^{21}$
            & $250$--$1000$
            & $550$
            & $17.679$
            & $2.462 + 3.459C_{22}(\cos\Psi-0.663) + 0.259J_2$
            & $0.0090$
            & $0.98$ \\
        \hline
    \end{tabular}
\caption{Physical parameters and best-fit disruption-limit equations for Chiron, Chariklo, Haumea, and Quaoar. In all cases, the primary and satellite have the same bulk density ($\rho_p=\rho$). The first term approximately corresponds to the classical Roche limit, while the second and third terms represent corrections associated with the primary's equatorial asymmetry and oblateness, respectively. The last two columns give the root mean square error (RMSE) on the test set and the coefficient of determination ($R^2$). The physical parameters were obtained from \cite{Braga2014,Ortiz2017,Morgado2023,Pereira2023,Proudfoot2025,Proudfoot2025b,Pereira2025}.}
\label{Tab_equations}
\end{table*}

All equations converge to very similar results, with the constants in each term generally differing only in the third decimal place. 
These small variations can be attributed to the fitting errors, which are $<0.0099$. The main exception is the disruption-limit equation fitted for Chariklo, and this difference arises because the symbolic regression model identified a $\cos^2(\Psi)$ term for Chariklo, whereas the equations for the other bodies depend on $\cos(\Psi)$. Despite this difference, the overall structure remains consistent across all four equations.

Table~\ref{SR_equation} shows the Hall of Fame of equations fitted to the dataset containing systems with different primary and satellite physical properties, ranked by increasing complexity. The simplest equation (N1) corresponds to the median value, while equation N2 resembles the classical Roche limit, $2.514(\rho_p/\rho)^{1/3}$. In equation N3, the algorithm introduces a constant correction to the classical Roche limit, and subsequent equations progressively incorporate additional terms involving $\cos(\Psi)$. The relevance of this angular dependence is consistent with the findings of \citet{Holsapple2006,Holsapple2008}, who showed that the orientation of the satellite's tidal bulge affects its equilibrium figure and disruption limit. 

The coefficient $C_{22}$ is introduced in equation N5, while $J_2$ appears in equation N8. In equation N9, the correction to the classical Roche limit also acquires an explicit dependence on the density ratio, consistent with the trends discussed in Section~\ref{sec_density}. Further increases in equation complexity do not yield significant improvements in performance.

\begin{table*}[h!]
\begin{center}
\begin{tabular}{clcc}
    \hline
    Nº & Disruption limit equation & RMSE & $R^2$ \\
    \hline
    
    N1 & $2.562$ & 0.4448 & $-0.0010$ \\
    
    N2 & $2.514\left(\dfrac{\rho_p}{\rho}\right)^{1/3}$ & 0.0869 & 0.9618 \\
    
    N3 & $2.414\left(\dfrac{\rho_p}{\rho}\right)^{1/3}+0.105$ & 0.0853 & 0.9632 \\
    
    N4 & $2.393\left(\dfrac{\rho_p}{\rho}\right)^{1/3}+0.199\cos\Psi$ & 0.0556 & 0.9843 \\
    
    N5 & $2.498\left(\dfrac{\rho_p}{\rho}\right)^{1/3}+C_{22}(2.498\cos\Psi-1.304)$ & 0.0374 & 0.9929 \\
    
    N6 & $2.496\left(\dfrac{\rho_p}{\rho}\right)^{1/3}+3.513C_{22}(\cos\Psi-0.549)$ & 0.0293 & 0.9957 \\
    
    N7 & $2.415\left(\dfrac{\rho_p}{\rho}\right)^{1/3}+3.518C_{22}(\cos\Psi-0.573)+0.091$ & 0.0254 & 0.9967 \\
    
    N8 & $2.454\left(\dfrac{\rho_p}{\rho}\right)^{1/3}+3.512C_{22}(\cos\Psi-0.680)+0.297J_2$ & 0.0174 & 0.9985 \\
    
    \textbf{N9} & ${\bf 2.465\left(\dfrac{\rho_p}{\rho}\right)^{1/3}+\left(4.675-\dfrac{\rho_p}{\rho}\right)\left[C_{22}(\cos\Psi-0.675)+0.072J_2\right]}$ & \textbf{0.0120} & \textbf{0.9993} \\
    
    N10 & $2.464\left(\dfrac{\rho_p}{\rho}\right)^{1/3}+\left[3.979-\dfrac{\rho_p}{\rho}+\cos\Psi\right]\left[C_{22}(\cos\Psi-0.701)+0.074J_2\right]$ & 0.0104 & 0.9995 \\
    
    N11 & $2.405\left(\dfrac{\rho_p}{\rho}\right)^{1/3}+C_{22}(\cos\Psi-0.696)\left(\cos\Psi+4.250-\dfrac{\rho_p}{\rho}\right)+0.258J_2+0.060$ & 0.0097 & 0.9995 \\
    
    N12 & $2.406\left(\dfrac{\rho_p}{\rho}\right)^{1/3}+C_{22}(\cos\Psi-0.697)\left[\cos\Psi+5.583-2.458\left(\dfrac{\rho_p}{\rho}\right)^{1/3}\right]+0.260J_2+0.058$ & 0.0086 & 0.9996 \\
    
    N13 & $2.408\left(\dfrac{\rho_p}{\rho}\right)^{1/3}+C_{22}(\cos\Psi-0.697)\left(\cos\Psi+4.481-2.514\left(\dfrac{\rho_p}{\rho}\right)^{1/3}\right)+0.259J_2+0.057$ & 0.0086 & 0.9996 \\
    
    N14 & $2.413\left(\dfrac{\rho_p}{\rho}\right)^{1/3}+C_{22}\left\{C_{22}\left[(\cos\Psi-0.659)\left(\cos\Psi+4.427-2.521\left(\dfrac{\rho_p}{\rho}\right)^{1/3}\right)-1\right]+0.258J_2+0.049\right\}$ & 0.0083 & 0.9996 \\
    
    \hline
\end{tabular}
\caption{Hall of Fame of the symbolic regression models, presenting the best-performing equations ranked by complexity. The first column identifies each equation, the second presents the disruption-limit equation expressed in units of the primary's radius, and the third and fourth columns give the RMSE and $R^2$ values, respectively. Equation 9 (in bold) provides the best fit, beyond which further increases in model complexity do not yield significant improvements in performance.}
\label{SR_equation}
\end{center}
\end{table*}

According to the ranking presented in Table~\ref{SR_equation}, the parameters influencing the disruption limit, in order of importance are the density ratio between the primary and satellite bodies, the relative orientation angle $\Psi$, and the coefficients $C_{22}$ and $J_2$. Our preferred expression for the disruption limit, derived from the symbolic regression analysis and expressed in units of the primary's radius, is
\begin{equation}
    R_{disruption} = 2.465\left(\dfrac{\rho_p}{\rho}\right)^{1/3}+\left(4.675-\dfrac{\rho_p}{\rho}\right)\left[C_{22}(\cos\Psi-0.675)+0.072J_2\right] \pm 0.012 
\label{Eq_disruption}
\end{equation}

When the coefficients $C_{22}$ and $J_2$ are set to zero, Equation~\ref{Eq_disruption} approximately reduces to the classical Roche limit. For the case in which the primary and satellite have equal densities, the equation simplifies to $ R_{\mathrm{disruption}}=2.465 + 3.675C_{22}(\cos\Psi - 0.675) + 0.264J_2 $, which is consistent with the equations presented in Table~\ref{Tab_equations}. These disruption-limit equations exhibit similar functional forms and coefficients, suggesting that the expression is not strongly dependent on the specific system considered, particularly the primary's mass and rotation period.

\section{Implications for small bodies bearing rings and conclusions} \label{sec_smallbodies}
\label{sec_implications}
\begin{figure*}
    \fig{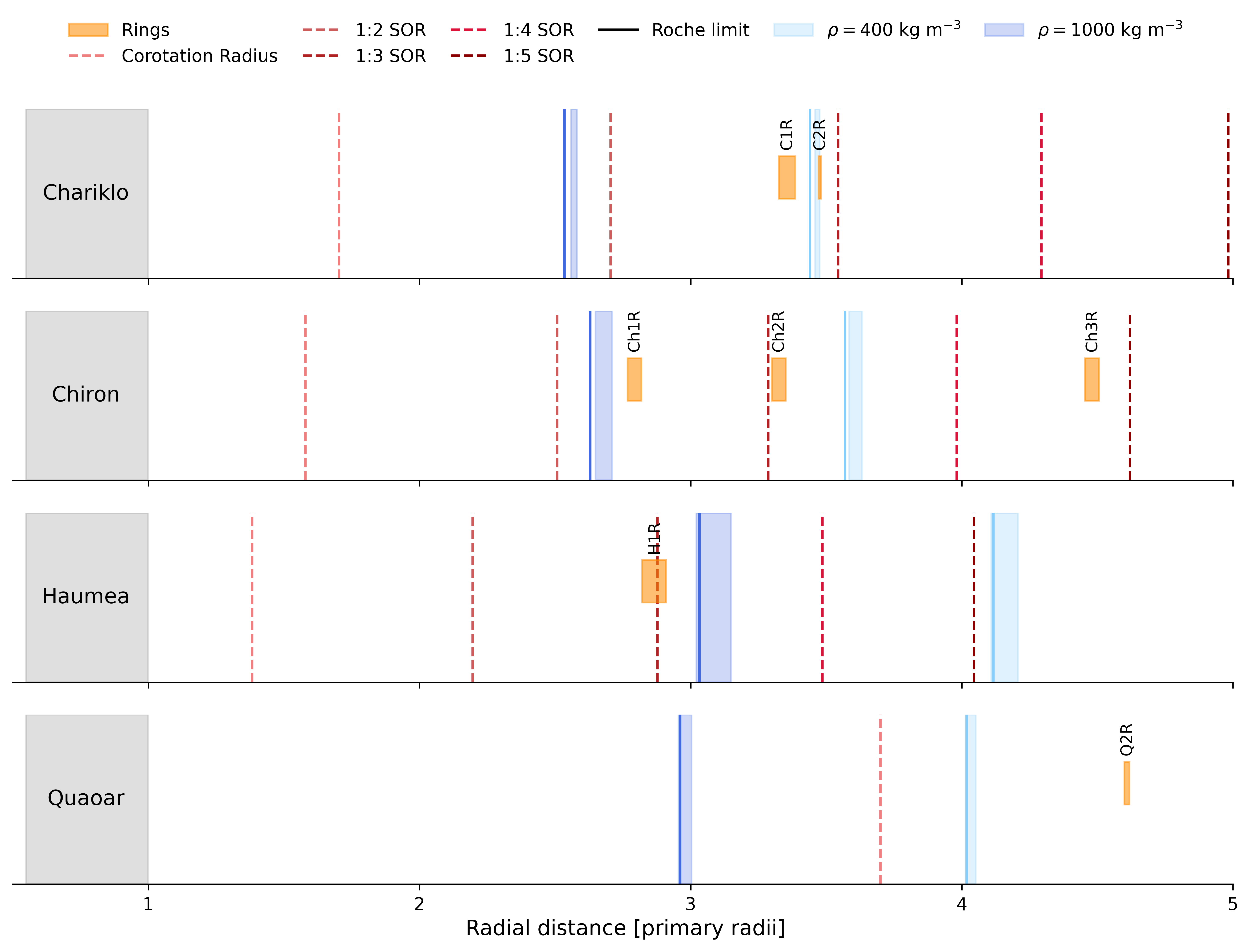}{1.0\textwidth}{}
\caption{Normalized radial locations of the rings, disruption limits, and spin-orbit resonances (SORs) for the Chariklo, Chiron, Haumea, and Quaoar systems. The grey shaded region marks the central body, while the orange vertical bands indicate the observed radial extent of the rings. The dashed vertical lines show the corotation radius and the theoretical SOR locations. The vertical solid lines mark the classical Roche limit, while the shaded regions show the radial interval spanned by the minimum and maximum disruption limits obtained for different values of $\Psi$. Both quantities are computed for two different bulk densities of the ring particles: $\rho=400~{\rm kg\,m^{-3}}$ (light blue) and $\rho=1000~{\rm kg\,m^{-3}}$ (blue). The former represents porous icy particles, while the latter corresponds to compact icy particles. The physical and orbital parameters of the Quaoar and Chiron systems were taken from \citet{Morgado2023,Pereira2023,Proudfoot2025,Proudfoot2025b} and \citet{Pereira2025}, respectively, while those of Chariklo and Haumea were taken from \citet{Braga2014} and \citet{Ortiz2017}, respectively.}
  \label{fig:rings}
\end{figure*}

Figure~\ref{fig:rings} summarizes the dynamical environment around Chariklo, Chiron, Haumea, and Quaoar, showing the disruption regions computed here. In the specific case of ring-bearing minor bodies, their triaxial shapes give rise to spin-orbit resonances (SORs), which occur when a particle's orbital period is commensurate with the primary's spin period. These resonances can increase the eccentricities of orbiting material and promote radial confinement \citep[e.g.,][]{Ribeiro2023,Madeira2025,Madeira2026,Salo2026}. Coincidentally or not, the rings of these objects are observed to lie close to SORs, suggesting that these resonances may play a key role in maintaining the material in a particulate state, regardless of its location relative to the Roche limit \citep{Sicardy2025,Celletti2026,Madeira2026,Beauge2026}. We therefore also indicate the locations of the main SORs in Figure~\ref{fig:rings}. 

In all four systems, the minimum disruption radius is shifted outward relative to the Roche limit for a given particle bulk density. Nevertheless, this outward displacement does not alter the overall picture for Chiron, Haumea, and Quaoar: rings located inside (outside) the classical Roche limit remain inside (outside) the disruption region for realistic values of the particle bulk density. Chariklo is the only case for which this classification changes, as C2R lies outside the classical Roche limit for $\rho=400~{\rm kg\,m^{-3}}$ but at the outer edge of the corresponding disruption region.

This result shows that the location of a ring with respect to the Roche or disruption limit should not be regarded as a complete explanation for its stability, nor for its long-term survival as a particulate structure, as is often implicitly assumed in the literature. Ring stability is a more complex problem, which depends not only on the ring location and composition, but also on its surface density, optical depth, particle-size distribution, collisional evolution, among other factors \citep{Sickafoose2024,Salo2026}. 

The disruption-limit location can nevertheless provide important insights into the possible origin of a ring system. The close proximity of Chariklo rings to the disruption region for $\rho\sim400~{\rm kg\,m^{-3}}$ may indicate that they formed from the tidal disruption of a porous icy satellite, whose porosity would account for such a low bulk density. In this scenario, the gap between the rings could be maintained by a sub-kilometre moonlet remnant of the disruption event \citep{Giuliatti2023}. This disrupted body could have been either a pre-existing satellite driven inward or to high eccentricity, or a newly formed satellite subsequently disrupted or stripped within a disk. 

A similar interpretation may apply to the H1R and Ch2R rings. Since these rings lie close to the 1:3 SOR and bit beyond the disruption region for compact icy material in the case of Haumea, and for porous icy material in the case of Chiron, they may correspond to material confined after the disruption of a satellite. This interpretation becomes particularly interesting given that our disruption region is computed for an inviscid fluid satellite, whereas a real object is expected to disrupt closer to the primary, depending on its material strength \citep{Holsapple2008,Madeira2023A}, and thus possibly closer to the 1:3 SOR. For Chiron, however, the frequent outbursts of the primary appear to provide a more natural origin for Ch2R and the other rings in the system \citep{Pan2016,Proudfoot2025a,Pereira2025}, although a scenario involving coagulation followed by disruption could have taken place.

Quaoar remains the most puzzling system, since the disruption region provides little insight into its formation. The presence of two rings, a newly discovered feature dynamically consistent with a satellite \citep{Nolthenius2025,Proudfoot2025a,Braga2026}, and possibly additional satellites \citep{Proudfoot2025a,Madeira2026} suggests that an extended disk may have once existed around Quaoar, giving rise to both rings and satellites. The slow rotation of the primary provides another possible indication of such a disk \citep{Regaly2025}: it may have resulted either from the impact that generated the disk or from the outward migration of satellites, which would have progressively slowed Quaoar's rotation, shifted the SOR locations outward, and possibly captured and transported the rings \citep{Beauge2025}.

The results presented here show that the triaxial shape of the primary affects the equilibrium figure of an inviscid fluid satellite, progressively shifting the disruption limit outward relative to the classical Roche-limit prediction as the primary's oblateness ($J_2$) and equatorial asymmetry ($C_{22}$) increase. For planetary-scale objects, their relatively low values of $J_{2}$ and $C_{22}$ make the classical Roche limit a good approximation, whereas for minor bodies the correction can become significant. In addition to the primary's shape, the disruption location depends on the orientation of its figure, defining a radial disruption region rather than a single critical distance for systems outside the double-synchronous configuration. The satellite inclination can also modify this location, although the effect becomes noticeable only at relatively high inclinations.

Our results, however, rely on the approximation of an inviscid fluid satellite. Real solid or granular satellites can sustain shear stresses, remaining stable at distances smaller than the fluid disruption region. Accounting for these effects would require introducing a material failure criterion, such as the Drucker-Prager criterion, together with information on the physical and mechanical properties of the satellite, including its internal friction and cohesive strength \citep{Holsapple2006,Holsapple2008}. Phobos provides a clear example, as it lies inside both the classical Roche limit and the disruption region computed using our methodology, while remaining intact due to its material strength \citep{Black2015}. The disruption regions derived here should therefore be interpreted as fluid limits rather than universal boundaries for the disruption of real satellites.

The correction to the classical Roche limit derived here is most relevant for Haumea and Chariklo among the known ring-bearing minor bodies. In the cases of Chiron and Quaoar, the rings that lie outside the classical Roche limit remain outside the disruption regions for realistic particle densities, requiring additional mechanisms to explain their persistence as particulate structures. Thus, the long-term maintenance of these rings appears to depend more broadly on particle properties, collisional evolution, and resonant effects driven by the non-axisymmetric gravity field of the primary, rather than on the location of the disruption limit alone.

\begin{acknowledgments}
We thank both reviewers for their insightful comments and suggestions, which helped us substantially improve the quality of this manuscript. We also thank the Brazilian government for its support of basic scientific research, and sincerely hope that such support will continue in the years to come. GM acknowledges the infrastructure provided by the Multiuser Data Processing Center of the National Observatory (CPDON). 
T.F.L.L. Pinheiro acknowledge support from the Conselho Nacional de Desenvolvimento Científico e Tecnológico (CNPq) - Proc. 313994/2025-0.
During the preparation of this work, the authors used ChatGPT-6 in order to improve readability. 
After using this tool, the authors reviewed and edited the content as needed and take full responsibility for the content of the publication.
This work was developed within the framework of the research project conducted by the Group of Origin and Dynamics of Satellites (\url{http://gusmadeira.github.io/gods}).
\end{acknowledgments}

\begin{contribution}
G.M. conceived the study, developed the numerical method, performed the calculations, analysed the results, and wrote the manuscript. T.P. performed the symbolic regression analysis and contributed to the writing of specific sections of the manuscript.
\end{contribution}

\bibliography{main}{}
\bibliographystyle{aasjournalv7}



\end{document}